# Emerging multidisciplinary research across database management systems

Anisoara Nica[1], Fabian M. Suchanek[2] [*] and Aparna S. Varde[3]
1. Sybase iAnywhere, Waterloo, ON, Canada
2. INRIA Saclay, Paris, France
3. Department of Computer Science, Montclair State University, Montclair, NJ, USA
anica@sybase.com, fabian@suchanek.name, vardea@montclair.edu

## ABSTRACT

The database community is exploring more and more multidisciplinary avenues: Data semantics overlaps with ontology management; reasoning tasks venture into the domain of artificial intelligence; and data stream management and information retrieval shake hands, e.g., when processing Web click-streams. These new research avenues become evident, for example, in the topics that doctoral students choose for their dissertations. This paper surveys the emerging multidisciplinary research by doctoral students in database systems and related areas. It is based on the PIKM 2010, which is the 3$^{rd}$ Ph.D. workshop at the International Conference on Information and Knowledge Management (CIKM). The topics addressed include ontology development, data streams, natural language processing, medical databases, green energy, cloud computing, and exploratory search. In addition to core ideas from the workshop, we list some open research questions in these multidisciplinary areas.

## 1. SURVEY OF TOPICS

Our survey groups the topics of the workshop into six research domains. We start with two core areas of database research, data mining and stream processing. We proceed to the sister domain of database management, Information Retrieval, before venturing into Information Extraction and the relatively new area of Privacy and Trust. We conclude with an application-driven domain, data warehouse management. We shed light on new research proposals in each of these domains, as well as on open issues.

### 1.1. Data mining

Data mining, the science of discovering meaningful knowledge in data, has always been a core area of database research. Nowadays, with the advent of large semantic knowledge bases, the area faces new challenges, for example with the task of integrating and aligning ontologies. The keynote of our workshop, *Leveraging Logical and Statistical Inference in Ontology Alignment* [9] by Professor Renée J. Miller from the University of Toronto, gave an overview on the work done in this area. It also presented a new approach, "**I**ntegrated **L**earning **I**n **A**lignment of **D**ata and **S**chema" (ILIADS), which combines data matching and logical reasoning to achieve better matching of ontologies. Professor Miller ended her talk with advice for graduate students based on her own experience – both as a student and as a graduate adviser.

Another problem in the area of data mining is frequent subgraph mining. Garcia et al. [5] study this problem with techniques available inside a database management system. Three fundamental research problems under a database approach are discussed: efficient graph storage and indexing, searching for frequent subgraphs and finding subgraph isomorphism using SQL. The solutions toward these problems are outlined, together with the preliminary experimental validation, focusing on query optimizations and time complexity.

Park et al. [13] research the issue of finding statistical correlations among pairs of attribute values in relational database systems. By extending the Bayesian network models the authors provide a probabilistic ranking function based on a limited assumption of value independence. Experimental results show that the proposed model improves the retrieval effectiveness on real datasets and has a reasonable query processing time compared to previous related work.

Discussions with the speakers at the workshop pointed to open research avenues in the area of data mining in general. One such avenue, inspired by the keynote talk, is investigating the usefulness of data mining in the Semantic Web: Data can be mined *from* the Semantic Web (for tasks such as summarization or alignment) as well as *for* the Semantic Web (with the purpose of adding new knowledge to an ontology). The interplay of these two directions appears still largely unexplored.

---

[*] The work by Fabian Suchanek has been partially funded by the European Research Council under the European Community's Seventh Framework Programme (FP7/2007-2013) / ERC grant Webdam, agreement 226513. http://webdam.inria.fr/

## 1.2. Stream processing

As more and more data (such as sensor data) becomes available in the form of continuous streams, new avenues of research open up. Amiguet et al. [2], for example, address the issue of semantic changes (annotations) in the stream data processing of sensor networks. The propagation of annotations through the workflow, as part of the data processing, raises interesting research questions related to how the propagation depends on the structural and temporal contributions, and on the annotation significance. The work shows its practical viability through a case study from a climate forecasting application that processes temperature sensor data. This paper won the best paper award in the PIKM 2010 workshop.

Another problem in stream processing is the adaptation of the system to different utilizations. Farag et al. [4] propose new algorithms for providing good data stream system performance during periods of peak load and periods of delays. The "External Memory Sliding Window Join" (EM-SWJoin) algorithm utilizes external memory data structures to adapt to the variable data arrival rates while keeping disk access latency at minimum. The "ADaptive Execution of DAta Streams" (ADEDAS) algorithm guarantees ordered release of output results to the user/application while controlling the impact of delays over stream processing. Thus, it addresses the problem of releasing output results in an increasing order of timestamp when the input data items arrive out-of-order. The authors exemplify their approach through a system that monitors online stocks.

In some scenarios, several data streams come together. Huq et al. [6] discuss how different data streams can be coordinated, how storage space can be optimized and how the database state can be kept consistent in a distributed scenario. The main application the authors have in mind is sensor-based experiments, for example based on motion sensors.

This last talk brought up the idea to investigate column-oriented data representations in data stream environments. Such a representation will require stronger coordination (horizontally in addition to vertically), but might potentially improve output efficiency and overall system performance.

## 1.3. Information retrieval

Recent years have seen an opening up of the border between the database research and information retrieval. Storing data and querying data go hand in hand, and this applies to both structured data and unstructured data. In many scenarios, queries can be classified into "lookup searches" and "exploratory searches". In lookup searches, users "look up" details on topics known to them; in exploratory searchers they "explore" new information. Mirizzi et al. [10] propose to combine these two concepts. They present a Web-based tool called "Lookup Discover Explore" (LED) that improves lookup search by adding exploration capabilities. LED supports a lookup step, an exploratory step and a meta search process. Given a keyword query, their system can display DBpedia resources with a graph explorer, a ranker and a context analyzer. One of the contributions of their work is a novel approach to browsing by exploiting the Semantic Web.

Di Buccio et al. [3] address the related problem of querying a document collection with short textual queries. Newer systems use relevance feedback from different sources of evidence. In this paper, the authors propose a uniform model for these evidence sources.

One area of Information Retrieval that is gaining more and more attention lately is the exploitation of the deep Web. Tjin-Kam-Jet [17] proposes to address this challenge in a distributed environment. Given the fact that the deep Web is up to two orders of magnitude larger than the surface Web, they argue that distribution might be the key to scalability. This work proposes to automatically convert free-text queries to structured queries for complex web forms in order to make the deep web more easily searchable. Challenges include developing a formal query description syntax, translating queries with correct interpretation, bridging the gap between user expectations and system capabilities, adapting query description for resource selection, ranking top k resources, merging results from resources to maximize precision, recall and suggestion ranking for users with respect to resources. He aims to evaluate this solution with a prototype system and user studies, criteria being processing time and user satisfaction.

Discussions following the talk [17] brought up the idea of studying distributed deep web searches in the context of handheld mobile devices. As mobile devices become practically ubiquitous (and reasonably powerful, too), it might become possible to utilize their capacities for distributed query processing.

## 1.4. Information extraction

Information extraction, in its widest sense, is the extraction of structured data from unstructured data. One domain where large corpora of unstructured text would particularly benefit from information extraction is the medical domain. Raghavan et al. [15] are concerned with extracting information from medical reports. A medical report is a natural language description of diagnoses, treatments or medications, together with structured information about the patient. The goal is to extract a chronology of events from the reports. Such chronologies can then be used to review a patient's history or to gather statistical data about the effectiveness or consequences of medical treatments. The task is challenging, because the reports use medical jargon and colloquial temporal expressions ("two days ago"). The paper conducts two initial case studies: In the first, machine learning on medical reports is used to determine whether patients qualify for Leukemia trials. In the second, a bio-specimen repository is augmented with data from medical reports. This additional data facilitates the classification of tissue probes and also information retrieval on the specimen database.

Kliegr [7] discusses the problem of noun phrase classification: Given a noun phrase (such as an image caption) and given a set of classes (such as "Nature" or "City"), the task is to assign a class to the noun phrase. This problem is challenging, because the classes are user-defined and the noun phrases do not necessarily have context. The paper proposes to map the noun phrase to a weighted set of related Wikipedia articles (a "bag of articles", BOA). Likewise, each class is mapped to a BOA. Then, the classification problem is reduced to determining the similarity between the noun phrase BOA and the class BOAs.

This last talk [7] has shown us that the possibilities that Wikipedia offers for information extraction are not yet fully exploited. The first talk [15] has pointed us to an area where Information Extraction (and possibly data mining) could find attractive applications: the medical domain.

## 1.5. Privacy and Trust

As more and more data is being produced, stored and made public, issues such as security, trust and privacy gain more and more attention. Consider for example pervasive health care systems. These are systems that monitor medical indicators on the patient and send their data to a medical center. These systems can greatly increase efficiency and safety, but come with the problem of data security. Acharya et al. [1] address the possible security threats in pervasive health care systems. One particular security loophole is the need to exchange encryption keys. The paper proposes a communication scheme that eliminates the need to broadcast encryption keys.

Security and trust are even more obvious issues when it comes to sharing data in cloud computing systems. Driven by this insight, Thorpe [16] develops a trust paradigm for the cloud. The paper considers interactions, reputation, knowledge and experience with respect to the cloud. Concepts such as user trust, cloud trust, cloud trust peer and cloud distrust are defined and an algorithm is proposed for a trust cloud context graph. The algorithm takes all the variables into account that are relevant to the cloud, i.e., machines, storage components and connected nodes. This encourages work on a context-neutral autonomous cloud monitor agent.

Just as [15], the authors of [1] target the medical domain. We take this as an indication that this domain can attract research from different areas of computer science, including security, privacy, data management, information extraction and systems research.

## 1.6. Workflow and Management

Research in databases can also touch quite practical issues such as the management of data centers. As data centers grow larger and larger and ever more powerful, pragmatic concerns such as heat management and environmental compatibility arise. Pawlish et al. [14], for example, are concerned with the environment-friendly management of data centers. The ultimate aim is to provide the data center manager with a tool that can help decide management questions, such as whether to buy new hardware and what hardware, in order to minimize the environmental impact of the facility. For this purpose, the paper first explains what data is necessary to support such decisions. These include, e.g., data on energy use, humidity, acoustic levels, and the carbon footprint of the center. For proof of concept, the authors have collected such data for one data center. Then, the paper proposes to build decision trees on top of these data and to use case based reasoning in order to determine economically sensible management steps. The paper outlines this framework supported by some real world examples.

Another issue in data warehousing environments is the problem of consistent data quality. As warehousing environments are often dynamically changing the quality of various resources and also the users' quality requirements, there is a need to capture quality changes on-the-fly and also provide automated quality notifications back to end users. Therefore, Li et al. [8] propose an extended data warehousing systems architecture that incorporates and extends the concepts of the "Quality Factory" (QF) and the "Quality Notification Service" (QNS), and adopts the "Data Warehouse Quality" (DWQ) methodology to provide both subjective and objective quality assessments of the end-users' quality requirements.

Apart from quality, efficiency of processes is also almost always an issue. Naseri et al. [11] are concerned with optimizing workflow management. They propose to systematically store, analyze and mine data about successful workflow executions in order to optimize workflow composition, workflow selection and workflow refinement.

Discussions during the workshop were particularly attracted to the concept of managing data warehouses. This area touches not just several domains of computer science (such as storage, querying, and quality estimation), but also other domains such as economics, the environmental sciences and the legal domain. The talks in this session have raised awareness for the practical impacts of our ever-growing amounts of data, their quality and their storage.

## 2. CONCLUSIONS

This paper presents a short survey of emerging research across database systems and related areas. It focused on the work that doctoral students presented at the PIKM workshop in ACM CIKM 2010. From a research point of view, the work is spread over areas as diverse as information extraction and workflow management, stream processing and security. From an application point of view, the work is concerned with cloud computing systems, sensor networks, practical aspects of data warehouse management and Web search. The medical domain with its applications has also attracted particular interest. The new research proposals as well as discussions with the students have illuminated some of the new research avenues in the domain of databases and beyond.

## 3. ACKNOWLEDGMENTS

We sincerely thank the CIKM 2010 organizers for their help and support in organizing the Ph.D. workshop PIKM. In particular we thank the General Chair, Jimmy Huang and the Workshops Chair, Mounia Lalmas, for their cooperation. We also express our sincere gratitude towards the 24 PC members of PIKM, spread over 12 countries across the globe. They comprised well-known experts from industry and academia and their exhaustive reviews helped to make this workshop a fruitful and enriching experience. All of this helped set the stage for writing this article.

## 4. REFERENCES

[1] D. Acharya and V. Kumar. A secure pervasive health care system using location dependent unicast key generation scheme. In PIKM 2010.

[2] J. Amiguet, A. Wombacher, and T. E. Klifman. Annotations: Dynamic semantics in stream processing. In PIKM 2010.

[3] E. Di Buccio and M. Melucci. Toward the design of a methodology to predict relevance through multiple sources of evidence. In PIKM 2010.

[4] F. Farag, M. Hammad, and R. Alhajj. Adaptive query processing in data stream management systems under limited memory resources. In PIKM 2010.

[5] W. Garcia, C. Ordonez, K. Zhao, and P. Chen. Efficient algorithms based on relational queries to mine frequent graphs. In PIKM 2010.

[6] M. R. Huq, A. Wombacher, and P. M. G. Apers. Identifying the challenges for optimizing the process to achieve reproducible results in E-science applications. In PIKM 2010.

[7] T. Kliegr. Entity classification by bag of Wikipedia articles. In PIKM 2010.

[8] Y. Li and K. M. Osei-Bryson. Quality factory and quality notification service in data warehouse. In PIKM 2010.

[9] R. J. Miller, Leveraging Logical and Statistical Inference in Ontology Alignment, Keynote Talk In PIKM 2010.

[10] R. Mirizzi and T. Di Noia. From exploratory search to web search and back. In PIKM 2010.

[11] M. Naseri and S. A. Ludwig. A multi-functional architecture addressing workflow and service challenges using provenance data. In PIKM 2010.

[12] A. Nica and A. S. Varde, editors. Proceedings of the Third Ph.D. Workshop in CIKM, PIKM 2010, Nineteenth ACM Conference on Information and Knowledge Management, CIKM 2010, Toronto, Canada, Oct. 2010. ACM.

[13] J. Park and S. G. Lee. Probabilistic ranking for relational databases based on correlations. In PIKM 2010.

[14] M. J. Pawlish and A. S. Varde. A decision support system for green data centers. In PIKM 2010.

[15] P. Raghavan and A. M. Lai. Leveraging natural language processing of clinical narratives for phenotype modeling. In PIKM 2010.

[16] S. Thorpe. Modeling a trust cloud context. In PIKM 2010.

[17] K.T. Tjin-Kam-Jet. Research proposal for distributed deep web search. In PIKM 2010.